\documentclass[conference]{IEEEtran}

\usepackage{graphicx}
\usepackage{amssymb}
\usepackage{amsmath}
\usepackage{calc}
\usepackage{array}
\usepackage{optidef}
\usepackage{subcaption}

\usepackage{cite}
\usepackage{comment}
\usepackage{algorithm}
\usepackage{algpseudocode}
\usepackage{caption}
\makeatletter

\newcommand{\Rmnum}[1]{\expandafter\@slowromancap\romannumeral #1@}
\makeatother
\begin{document}
\title{A Deep Learning Approach for Automotive Radar Interference Mitigation}

\author{Jiwoo Mun, Heasung Kim, and Jungwoo Lee \\
Department of Electrical and Computer Engineering, Seoul National University, Seoul, Korea \\
E-mail: \{jwmun, heasung1130\}@cml.snu.ac.kr, junglee@snu.ac.kr \vspace{-3mm}
}

\maketitle

\begin{abstract}
In automotive systems, a radar is a key component of autonomous driving. Using transmit and reflected radar signal by a target, we can capture the target range and velocity. However, when interference signals exist, noise floor increases and it severely affects the detectability of target objects. For these reasons, previous studies have been proposed to cancel interference or reconstruct original signals. However, the conventional signal processing methods for canceling the interference or reconstructing the transmit signals are difficult tasks, and also have many restrictions. In this work, we propose a novel approach to mitigate interference using deep learning. The proposed method provides high performance in various interference conditions and has low processing time. Moreover, we show that our proposed method achieves better performance compared to existing signal processing methods.
\end{abstract}


\section{Introduction}
\label{sec_int}
Radars mounted on advanced vehicles, such as autonomous vehicles, require a variety of functions, including detection of multi-target and long-range sensing. These functions must be performed accurately ensure user safety and solve collision problem between vehicles. Recent popular radar technologies include Frequency Modulated Continuous Wave (FMCW) or Chirp Sequence (CS) radars \cite{kronauge2014new,winkler2007range,stove1992linear}. However, it is difficult to perform the above functions with interference \cite{brooker2007mutual,goppelt2011analytical}.

Several techniques have been proposed to solve the problems related to interference \cite{watanabe2007interference,kunert2012eu,bechter2015automotive,barjenbruch2015method,wagner2018threshold}. \cite{watanabe2007interference} used the characteristics of the interference region in the time domain to remove the interference. \cite{bechter2015automotive} proposed a method of estimating the amplitude and frequency of the interference signal to recover the original signal as well as the interference elimination with high computational complexity. The paper \cite{wagner2018threshold} proposed an algorithm that requires a small computational complexity and showed that it detects targets within small distances without defining an adaptive threshold. The effect of interference still remains, however, because the target is not well detected when the interference signal source is closer to the radar than the target.

To the best of our knowledge, we are the first to  use a deep learning method to mitigate interference in time domain. Recently, the development of deep learning has been remarkable, and in particular, it has made significant achievements in image and language processing. Besides, these deep learning techniques have shown outstanding results in the field of signals, and \cite{yao2017deepsense} and \cite{yu2011deep} showed that deep learning can be useful in signal processing. Especially we apply the Recurrent Neural Network (RNN) model with Gated Recurrent Unit (GRU)\cite{chung2014empirical}, which is known to be suitable for processing sequence data, to remove interference and reconstruct transmit signal simultaneously. We can reconstruct transmit signal even in the presence of various interference signals, and the reconstructed signal can be used to detect objects through Fast Fourier Transform (FFT). In particular, through the learned network, signal processing can be done only with the matrix calculation, not with any iteration structure. Also, the algorithm does not require any adaptive threshold. We show that our algorithm outperforms existing algorithms in experiments where noise and interference coexist.

The rest of this paper is organized as follows.
In Section \ref{sec_sys}, we introduce the system model considered in the paper. 
In Section \ref{sec_pro}, we show the deep learning model for our proposed algorithm.
In Section \ref{sec_num}, we show the simulation results for the proposed scheme.
Lastly, in Section \ref{sec_con}, we conclude this paper.
\section{System Model}
\label{sec_sys}
\subsection{CS Radar System}
\begin{figure}[ht]
\vspace{0mm}
 		\includegraphics[width=1\linewidth]{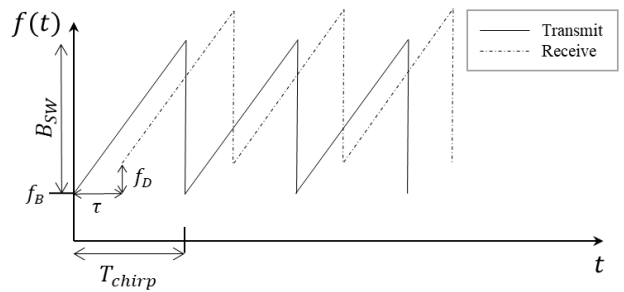}
 		\caption{CS waveform of transmit and received signal}
     	\label{chirp_signal_figure}
\end{figure}

\begin{figure}[ht]
\vspace{0mm}
 		\includegraphics[width=1\linewidth]{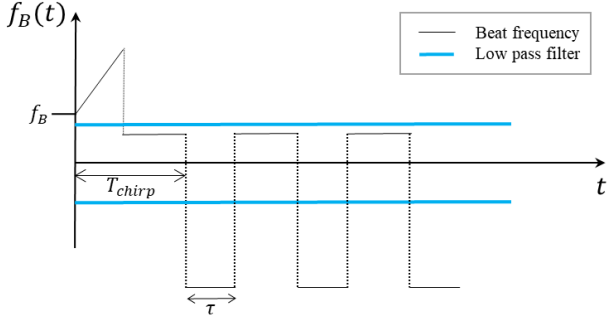}
 		\caption{Beat frequency}
     	\label{b_frequency_figure}
\end{figure}

One of the main radar waveforms is the CS waveform \cite{kronauge2014new,stove1992linear} as shown in Fig. \ref{chirp_signal_figure}.
If the transmit signal consists of $k$ linear frequency chirps, frequency and phase of the transmit signal are as follows.
\begin{equation}\label{eq1}
\begin{split}
f(t) &= f_{B} + \alpha (t-k T_{chirp}) \\
\phi(t) &= 2\pi \int_{0}^{t} f(t) dt\\ 
&= 2\pi (f_{B}t + \frac{1}{2} \alpha t^2 - \alpha k T_{chirp}t),
\end{split}
\end{equation}
where $B_{SW}$ is sweep bandwidth, $T_{chirp}$ is chirp duration, $\alpha = B_{SW}/T_{chirp}$ is slope of the CS waveform, and $f_{B}$ is carrier frequency of the transmit signal. The beat frequency is the difference between the transmit frequency and the received frequency. The beat frequency $f_{B}(t)$ is represented as Fig. \ref{b_frequency_figure}. A Low Pass Filter (LPF) can remove signals with higher absolute frequency value. So the remaining beat phase through the LPF can be represented as

\begin{equation}\label{eq2}
\begin{split}
\phi_{B}(t) &=\phi(t)-\phi(t-\tau) \\  &= 2\pi f_{B}\tau -\pi\alpha(\tau^2-2\tau t_k)  \\  &\text{if } \tau < t < T_{chirp}.
\end{split}
\end{equation}
We denote target range, target velocity and speed of light as $R$, $v$, $c$, respectively, and the propagation delay can be represented as $\tau$. Substituting $\tau=\frac{2(R + vt)}{c}$ and $t=kT_{chirp} + t_{k}$ into equation (\ref{eq2}) (if $t$ is present in $k$-th chirp), (\ref{eq2}) can be approximated
\begin{equation}\label{eq3}
\begin{split}
\phi_{B}(t) &=2\pi f_{B}\left(\frac{2R}{c} + \frac{2vt}{c}\right) \\  & - \pi\alpha \left (\left (\frac{2R}{c} + \frac{2vt}{c} \right)^2 - 2\left (\frac{2R}{c} + \frac{2vt}{c} \right )t_k \right) \\ &\approx 2\pi \left( \frac{2R}{c}f_{B} + \frac{2v}{c}f_{B}k\right. T_{chirp} \\
&+ \left(\frac{2\alpha R}{c} + \frac{2v}{c}f_{B}\right)t_{k}\left. \vphantom{\int_1^2}\right).
\end{split}
\end{equation}
Applying sampling as $t=nT_{s}$, phase of the beat signal $\phi_{B}[n, k]$ is written as
\begin{equation}\label{eq4}
\begin{split}
\phi_{B}[n, k] & = 2\pi\left(\frac{2R}{c}f_{B} + \frac{2v}{c}f_{B}k T_{chirp} \right. \\ 
 &+ \left(\frac{2\alpha R}{c} + \frac{2v}{c}f_{B} \right) nT_{s}\left. \vphantom{\int_1^2}\right).
\end{split}
\end{equation}
Using two dimensional Fast Fourier Transform (FFT), we can obtain following two values $f_R$ and $f_D$,
\begin{equation}\label{eq5}
\begin{split}
f_R &= \frac{2\alpha R}{c} \\
f_D &= \frac{2v}{c}f_B T_{chirp}.
\end{split}
\end{equation}
Range $R$ and velocity $v$ can be obtain by $f_R$ and $f_D$.
\begin{figure}[ht]
 		\includegraphics[width=1\linewidth]{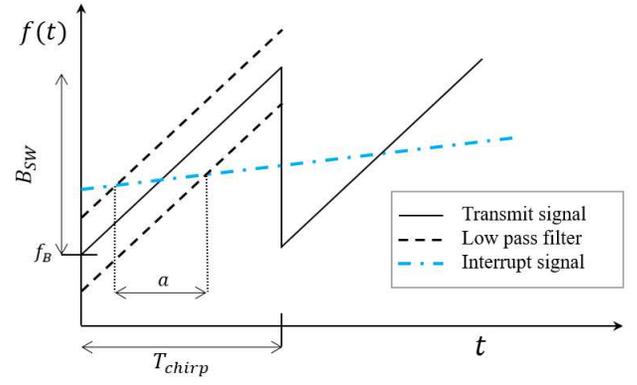}
 		\caption{Interrupted transmit signal, interference occurs in $a$.}
     	\label{interrupt_signal_figure}
\end{figure}

\begin{figure}
\hspace{7mm}
 		\includegraphics[width=0.8\linewidth]{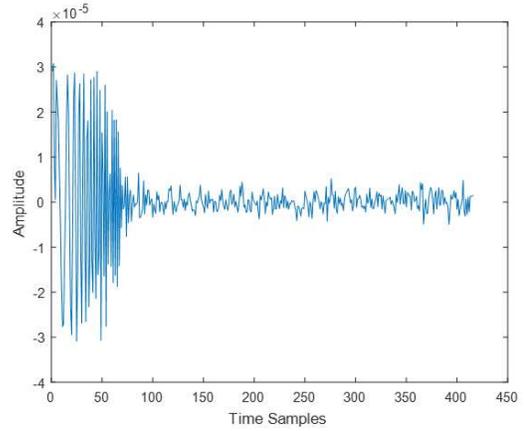}
 		\caption{Interrupted beat signal, interference occurs around the 0 to 80 samples.}
     	\label{interfered_beat_signal_figure}
\end{figure}
\subsection{Interrupted Radar Signal}
The equations in the previous subsection are derived in an ideal situation without interference. However, there will be a large error in distance and velocity estimation if interference occurs. In a typical driving situation, we usually encounter CS waveform signals, which have different slopes with the signal being sent, and interference situation would occur as shown in Fig. \ref{interrupt_signal_figure}. Since the beat frequency passes through the low pass filter, the interference occurs in the section $a$ only, not in the whole section. Fig. \ref{interfered_beat_signal_figure} shows that a large distortion occurs around 0 to 80 time samples, unlike the original beat signal. Conventionally, the interference is removed or the original beat signal is restored by using the characteristics of the time-domain beat signal. However, if noise and interference exist,  the cancellation of interference and the restoration of original beat signal are difficult with a traditional method.

\section{Interference mitigation using Deep learning}
In this section, we propose a deep neural network model which can be used for multi-interference mitigation without relying on adaptive threshold. 
\label{sec_pro}
\subsection{Deep Learning Model}
As shown in previous studies \cite{lipton2015critical}, RNN is known to be suitable for sequence data processing. Since the raw data before preprocessing is consecutive time samples, we apply RNN structure for interference cancellation and restoration in our model. Following equations represents the vanilla RNN elements.
\begin{equation}\label{eq6}
\begin{split}
h_t &= f_{W}(h_{t-1},x_t) \\ 
&= tanh(W_{hh}h_{t-1} + W_{xh}x_t) \\ 
y_t &= W_{hy}h_t.
\end{split}
\end{equation}
$x_t$ is the input vector, $h_t$ is the hidden state of the RNN network and $y_t$ is the output vector. $W_{hh}$ ,$W_{xh}$ and $W_{hy}$ are weight matrices of the hidden state to another hidden state, the input vector to the hidden state and the hidden state to the output vector, respectively. By using RNN, the network can learn the relation of consecutive samples.
\begin{figure}[ht]
\vspace{0mm}
 		\includegraphics[width=1\linewidth]{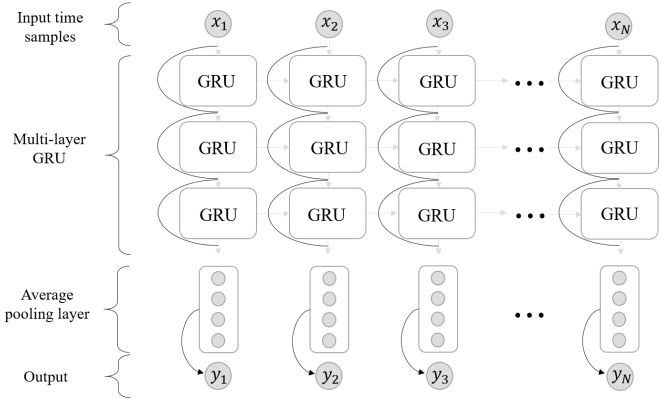}
 		\caption{Proposed deep learning model}
     	\label{deep_learning_model_figure}
\end{figure}
The input sequence may consist of hundreds of time samples. It may cause \textit{long-term dependency} problem in RNN \cite{bengio1994learning}. So we use a GRU cell to solve this problem in RNN. GRU has the same time series structure as RNN, but the contents of the cell are different. In the multi-layer GRU layer, each layer has a bidirectional structure, rather than one direction of the signal\cite{schuster1997bidirectional}. In addition, several GRU layers were piled up to learn various interference cases. The residual network\cite{he2016deep} is added between layers for better propagation of gradient flow. The residual connection is written as
\begin{equation}\label{eq7}
\begin{split}
X^{l+1} = X^l + GRU(X^l), (l=1,2,3, ... , L-1),
\end{split}
\end{equation}
where $X^l$ is $l$-th layer input vector of GRU cells, and $GRU(X^l)$ is $l$-th layer output vector of GRU cells. When the total time step is $N$ and the hidden state size is $H$, the output value of GRU network is $X^{L}\in\mathbb{R}^{H\times N}$. If we denote $x^{L}_i\in\mathbb{R}^{H}$ as the $i$th column vector of $X^{L}$, $X^L$ can be represented as $[x^L_1,x^L_2,...,x^L_N]$. To obtain the output dimension identical to the label dimension, we perform average pooling on $X^{L}$. The average pooling output $Y\in\mathbb{R}^{N}$ is written as
\begin{equation}\label{eq8}
\begin{split}
Y=[average(x^L_1),average(x^L_2),...,average(x^L_N)].
\end{split}
\end{equation}
To regularize the network, we applied drop out in each GRU Cells\cite{srivastava2014dropout}. The proposed RNN model is shown in Fig. \ref{deep_learning_model_figure}. 
\subsection{Optimizing Model}
The inputs is time-sampled interference beat signal, which is represented as $X^0 = X = [x_1,x_2,...,x_N]$, where $x_i\in\mathbb{R}$ is amplitude of beat signal$(i=1,...,N)$. Each input $X$ is normalized and satisfies the following equation. 
\begin{equation}\label{eq9}
\begin{split}
	\sum_{i=1}^N x_{i}^2=1.
\end{split}
\end{equation}
The output $Y$ is represented as $Y=[y_1,y_2,...,y_N]$, which has the same length as $X$. $\hat{Y}=[\hat{y_1},\hat{y_2},...,\hat{y_N}]$ is a beat signal with the same target condition as $X$ but without interference. We called $\hat{Y}$ as label. In order to minimize the difference between the two vectors $Y$ and $\hat{Y}$, the loss $L$ is defined as
\begin{equation}\label{eq10}
\begin{split}
L= \sum_{i=1}^N (\hat{y_i}-y_i)^2.
\end{split}
\end{equation}
The loss $L$ can be minimized by gradient descent. We use a gradient descent algorithm, Adam\cite{kingma2014adam}. As the training progresses, we get output $Y$ which is similar to label $\hat{Y}$. We can then use this value $Y$ to detect target range and velocity.
%
\section{Simulation Results}
\label{sec_num}
In this section, we introduce radar simulator parameters and deep learning model parameters. The proposed deep learning model is also compared with existing algorithms.
\begin{table}[ht]
\caption{Radar simulator random parameters}
\label{radar_simulator_parameter}
\begin{center}
\begin{tabular}{c||c||c}
\hline
Parameter & Min & Max \\
\hline
Center frequency & 76GHz & 78GHZ \\
Distance & 1m & 130m \\
Velocity & 0km/h & 50km/h  \\
Sweep bandwidth & 100MHz & 200MHz \\
Chirp duration & 20us & 40us \\
Target number & 1 & 2 \\
Interference number & 1 & 4 \\
\hline
\end{tabular}
\end{center}
\end{table}

\begin{table}[ht]
\caption{Deep learning hyperparameter}
\label{deep_learning_parameter}
\begin{center}
\begin{tabular}{c||c}
\hline
Hyperparameter & Value \\

\hline
Batch size & 128 \\
Learning rate & 1e-3 \\
Hidden layer size & 100 \\
Number of data & 150000 \\
Number of layer & 3 \\
Drop out rate & 0.3 \\
Optimizer & Adam \\
\hline
\end{tabular}
\end{center}
\end{table}
We have assumed a situation with multi-target, multi-interference, and Gaussian noise in order to reflect the practical situation. We use a randomly generated 150,000 time sampled input sequence (with interference) and 150,000 label sequence (no interference). The range of random parameters for training is shown in Table \ref{radar_simulator_parameter}. The transmit signal is the CS wave mentioned in Section \ref{sec_sys} and the interference waveform is the FMCW wave signal with different chirp slope (includes CS waveform, triangle sweep FMCW). The total number of chirps was 75 in both the desired and the interfering signals. The model proposed in Section \ref{sec_pro} is used and the hyperparameter used in the model is shown in Table \ref{deep_learning_parameter}. The deep learning model input and label are beat signals corresponding to one chirp of the transmit signal. To apply RNN, the input and label length must be constant. However, the number of samples of one chirp can vary depending on the sampling period of the signal. So we limit the maximum length of the input and label to 416 and cut the remaining part if the actual length is longer than that, and do zero-padding if it is smaller. In order to solve the \textit{exploding gradients} problem in the GRU structure, the gradient clipping method is used \cite{pascanu2013difficulty}.

\begin{figure}[ht] 
  \begin{subfigure}[b]{0.33\linewidth}
    \centering
    \includegraphics[width=0.99\linewidth]{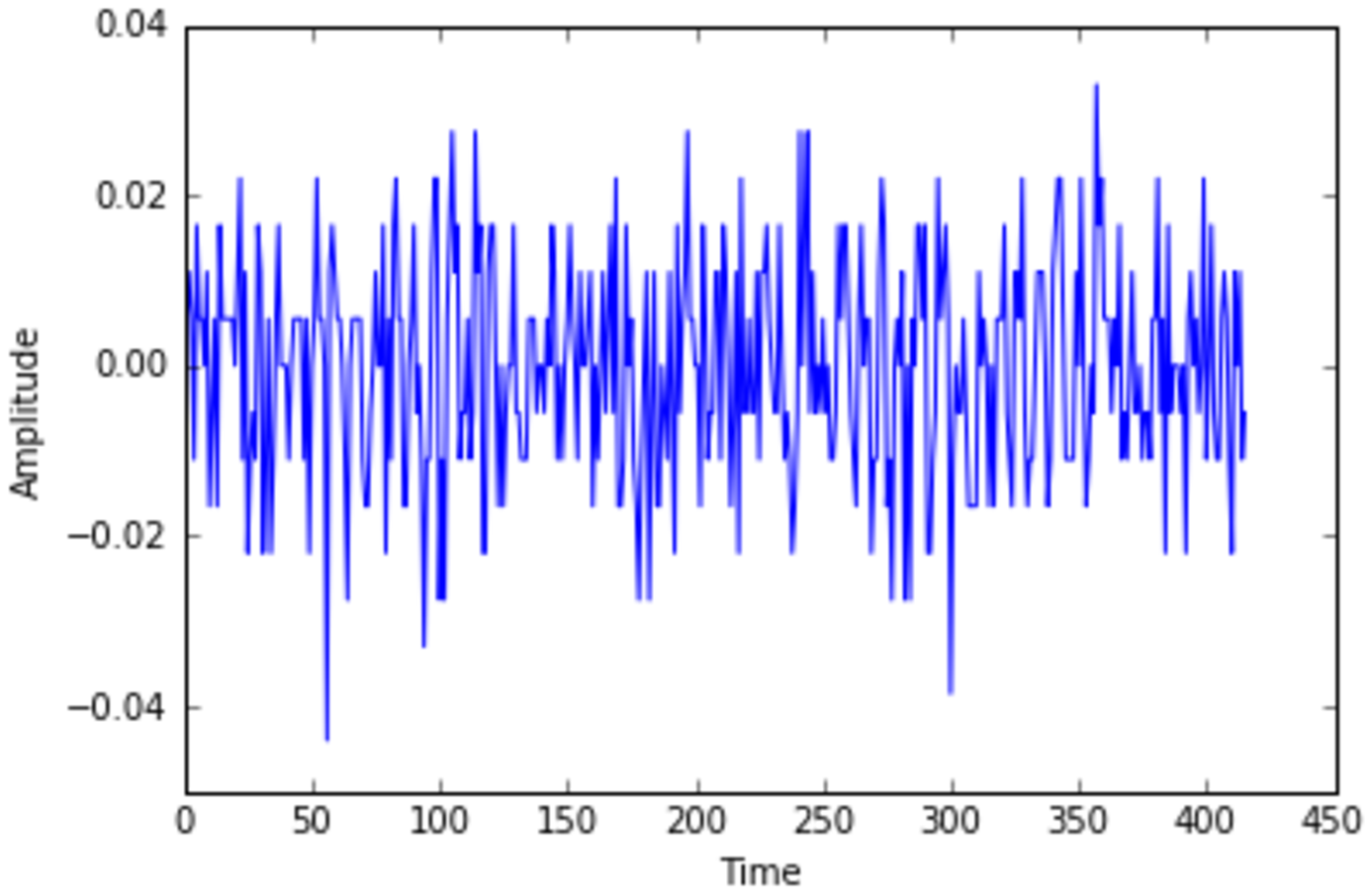} 
    \caption{Label} 
    \label{fig6:a} 
    \vspace{4ex}
  \end{subfigure}%
  \begin{subfigure}[b]{0.33\linewidth}
    \centering
    \includegraphics[width=0.99\linewidth]{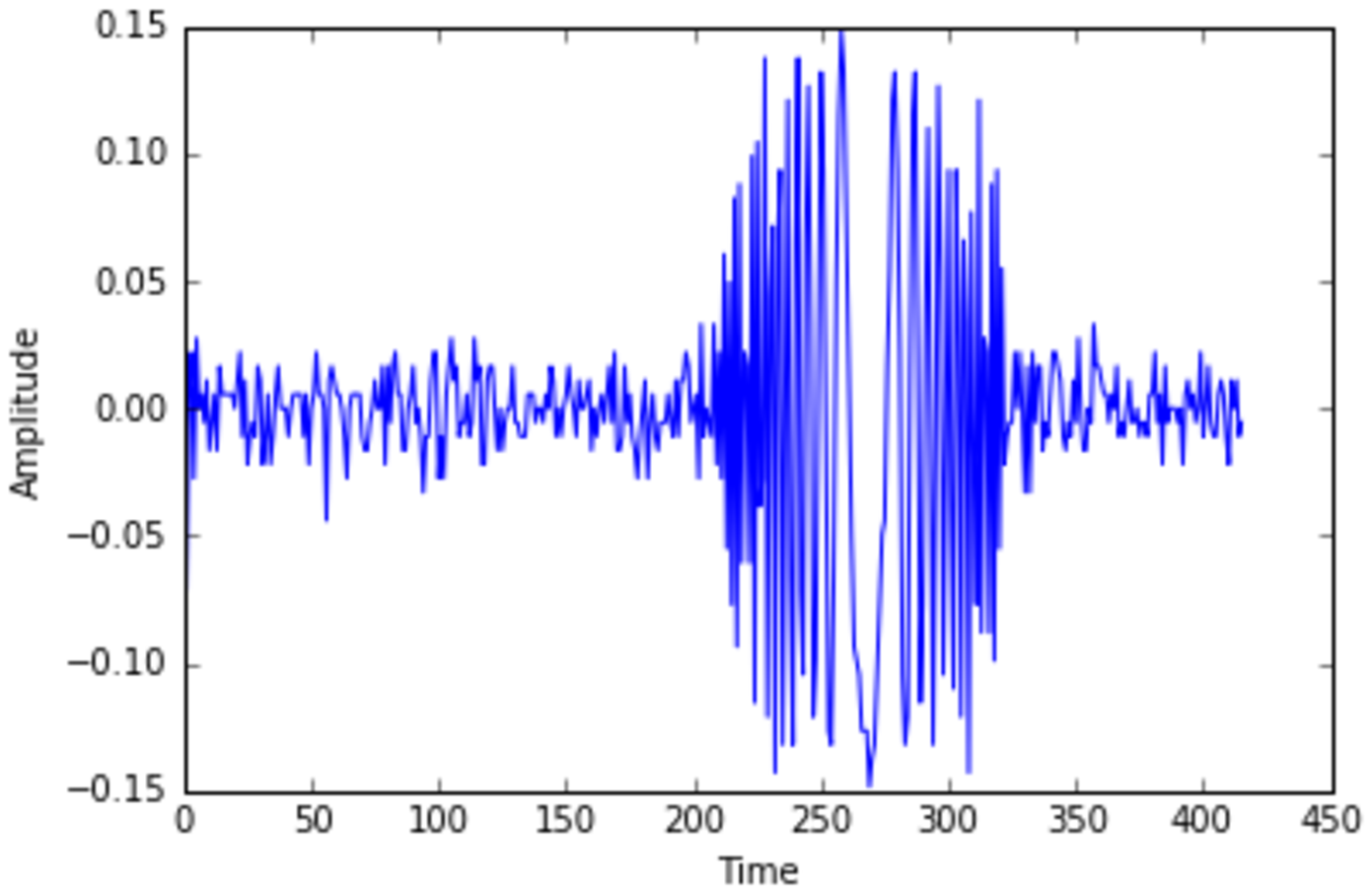} 
    \caption{Input} 
    \label{fig6:b} 
    \vspace{4ex}
  \end{subfigure}%
  \begin{subfigure}[b]{0.33\linewidth}
    \centering
    \includegraphics[width=0.99\linewidth]{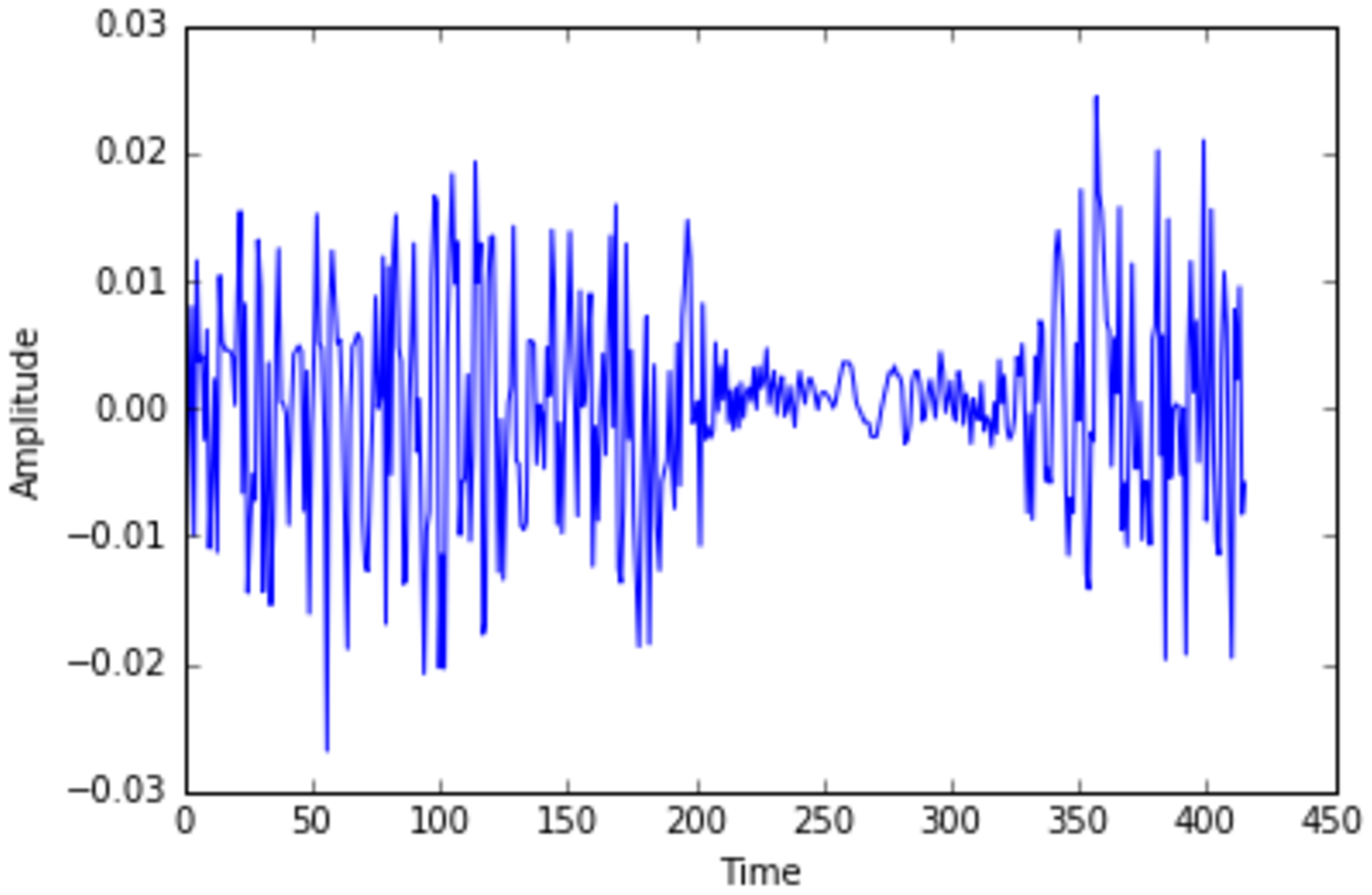} 
    \caption{Output} 
    \label{fig6:c} 
    \vspace{4ex}
  \end{subfigure}\\%
  \begin{subfigure}[b]{0.33\linewidth}
    \centering
    \includegraphics[width=0.99\linewidth]{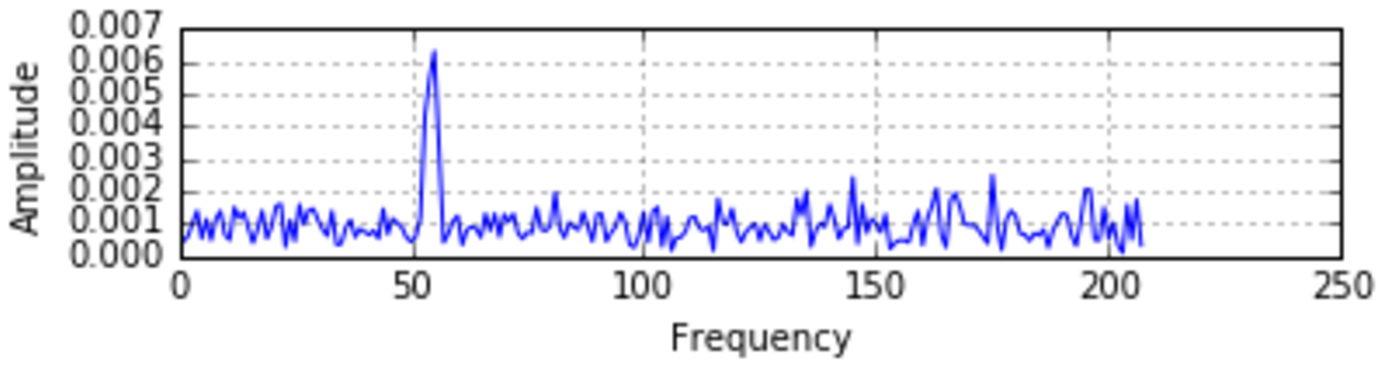} 
    \caption{FFT label} 
    \label{fig6:d} 
    \vspace{4ex}
  \end{subfigure}%
  \begin{subfigure}[b]{0.33\linewidth}
    \centering
    \includegraphics[width=0.99\linewidth]{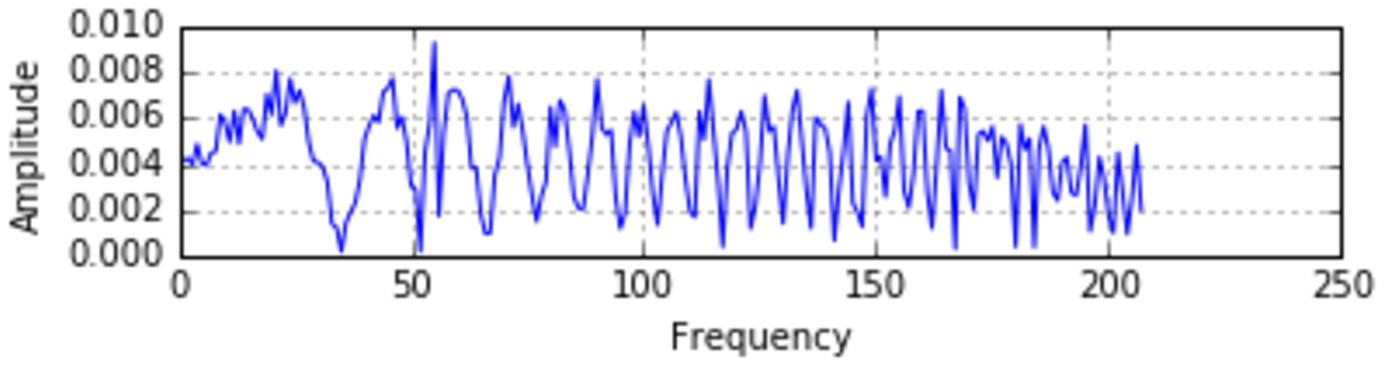} 
    \caption{FFT input} 
    \label{fig6:e} 
    \vspace{4ex}
  \end{subfigure}%
  \begin{subfigure}[b]{0.33\linewidth}
    \centering
    \includegraphics[width=0.99\linewidth]{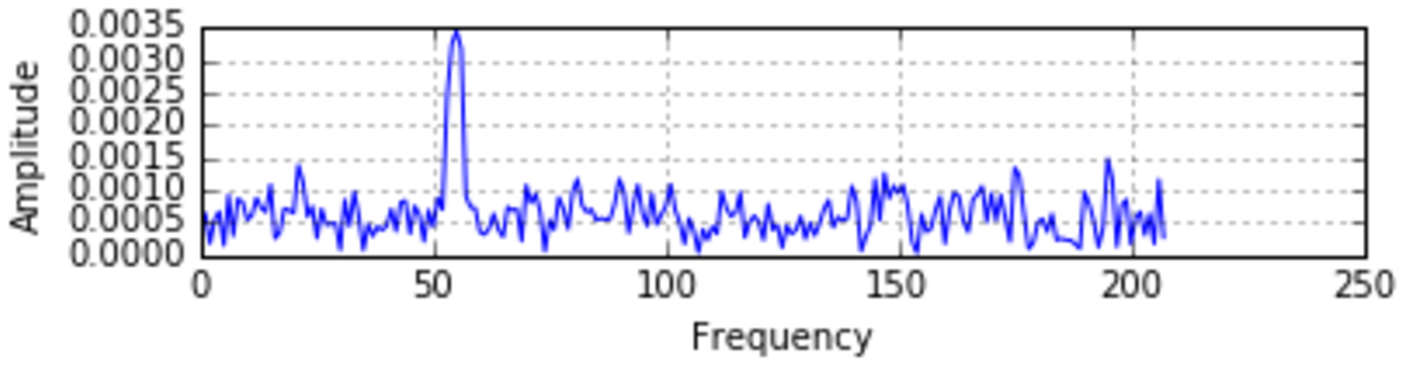} 
    \caption{FFT output} 
    \label{fig6:f} 
    \vspace{4ex}
  \end{subfigure}%
  \caption{Result of deep learning model. (a) to (c) is beat signal, (d) to (f) is FFT result of (a) to (c) signals respectively.}
  \label{fig6} 
\end{figure}

\begin{figure}[ht]
  \begin{subfigure}[b]{0.5\linewidth}
      \centering
      \includegraphics[width=0.99\linewidth]{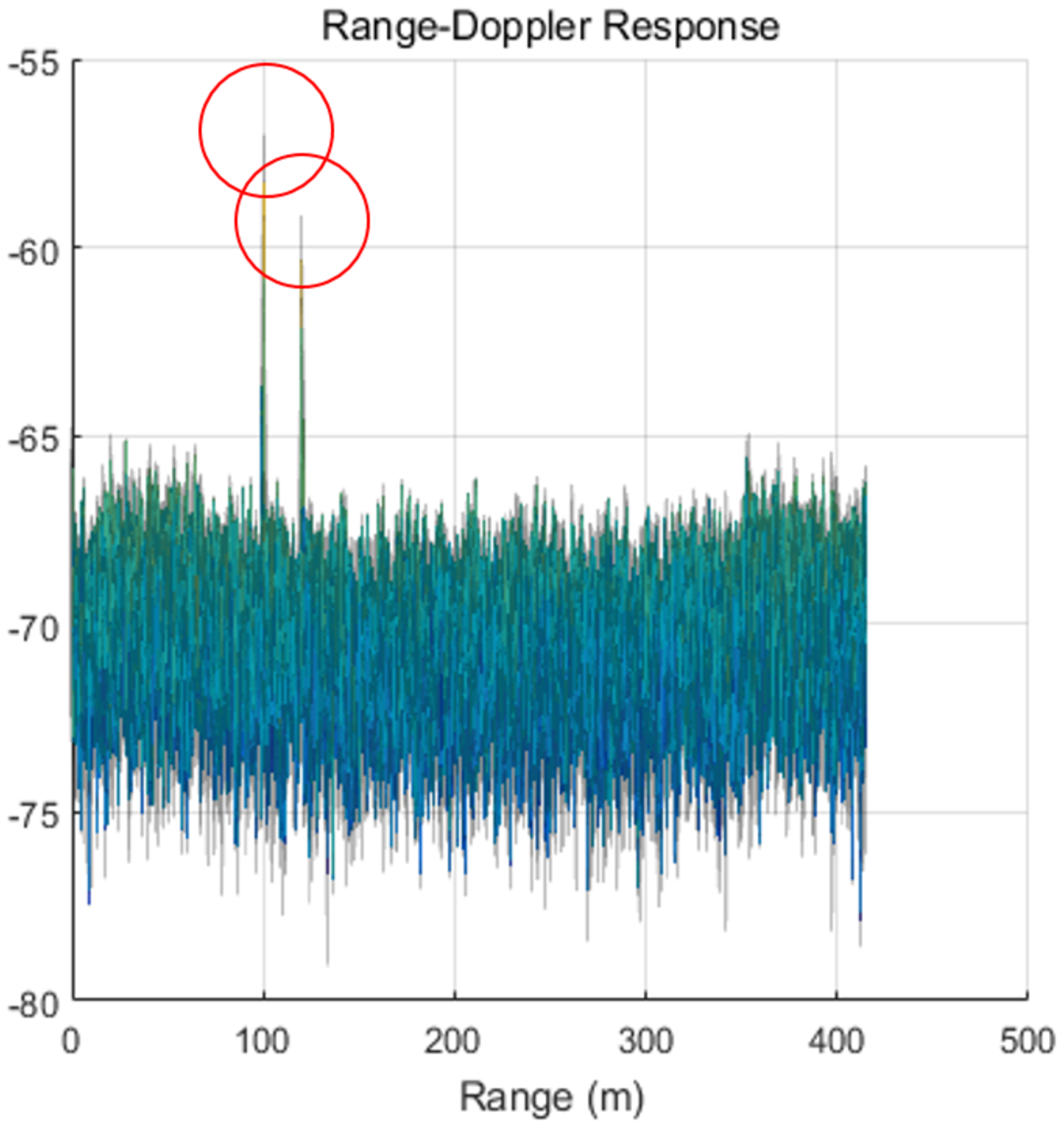} 
      \caption{Proposed} 
      \label{fig7:a} 
      \vspace{4ex}
    \end{subfigure}%
    \begin{subfigure}[b]{0.5\linewidth}
    \centering
    \includegraphics[width=0.99\linewidth]{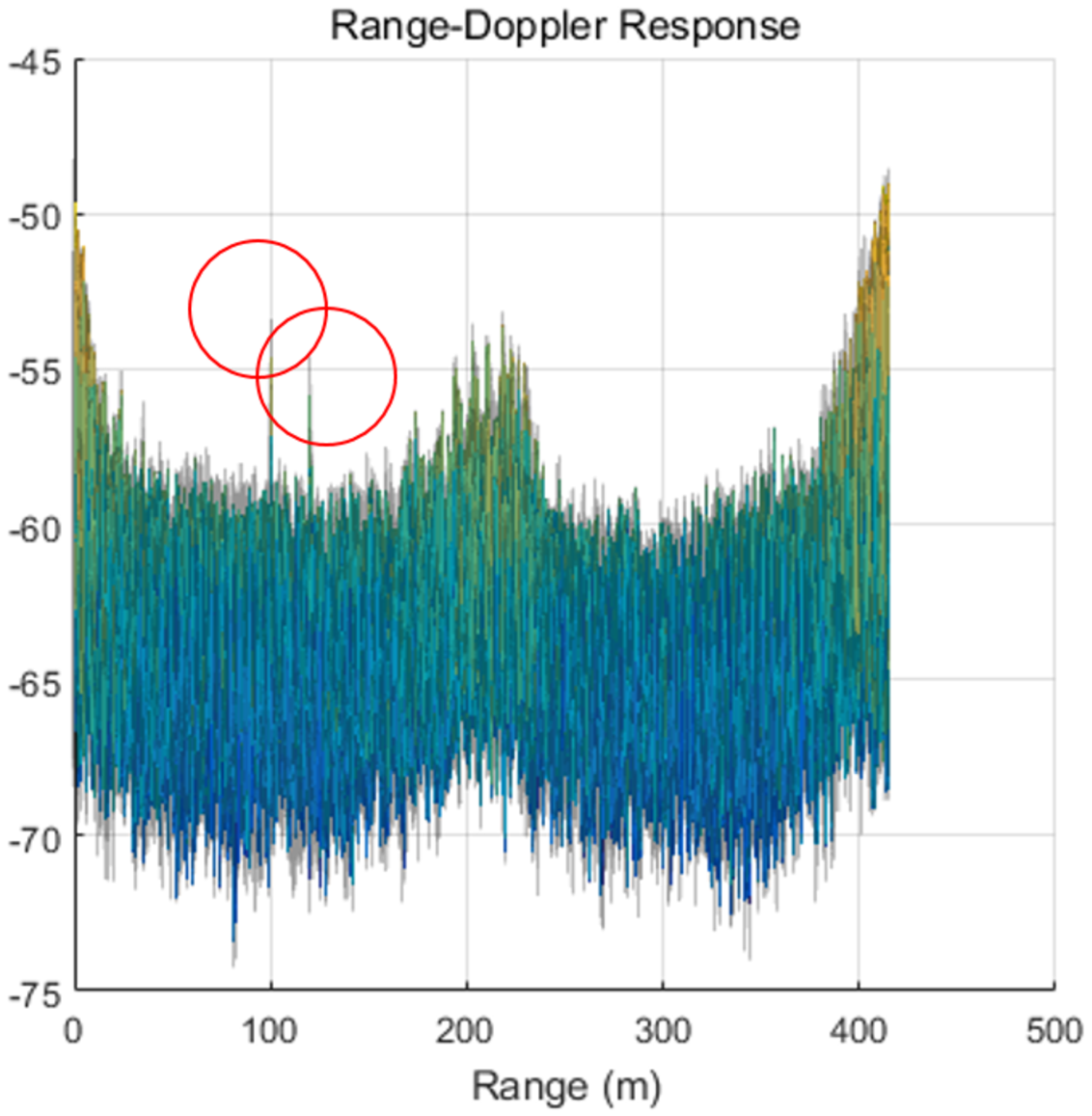} 
    \caption{Method \Rmnum{1}} 
    \label{fig7:b} 
    \vspace{4ex}
    \end{subfigure}\\%
    \begin{subfigure}[b]{0.5\linewidth}
    \centering
    \includegraphics[width=0.99\linewidth]{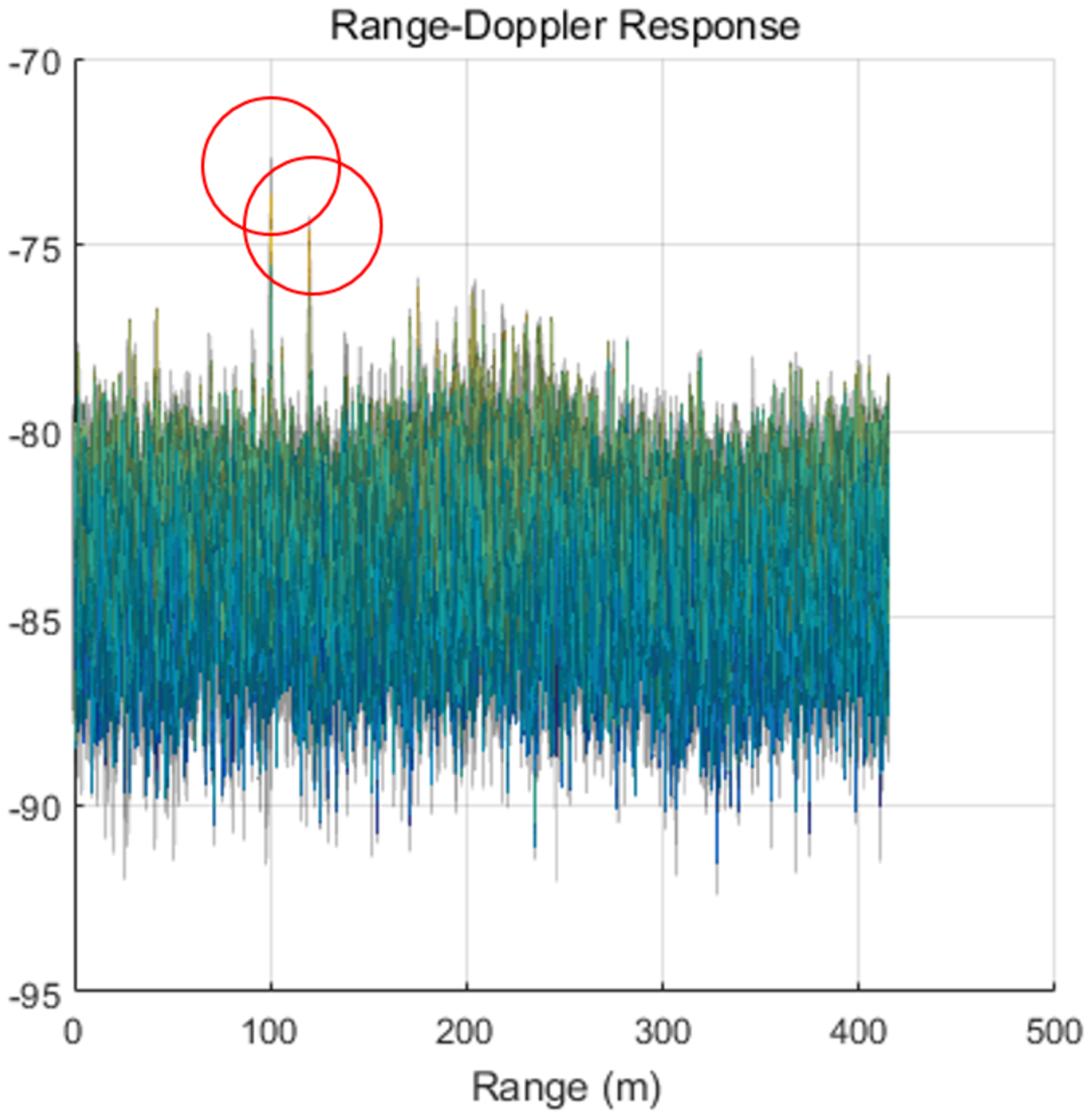} 
    \caption{Method \Rmnum{2}} 
    \label{fig7:c} 
    \vspace{4ex}
    \end{subfigure}%
    \begin{subfigure}[b]{0.5\linewidth}
    \centering
    \includegraphics[width=0.99\linewidth]{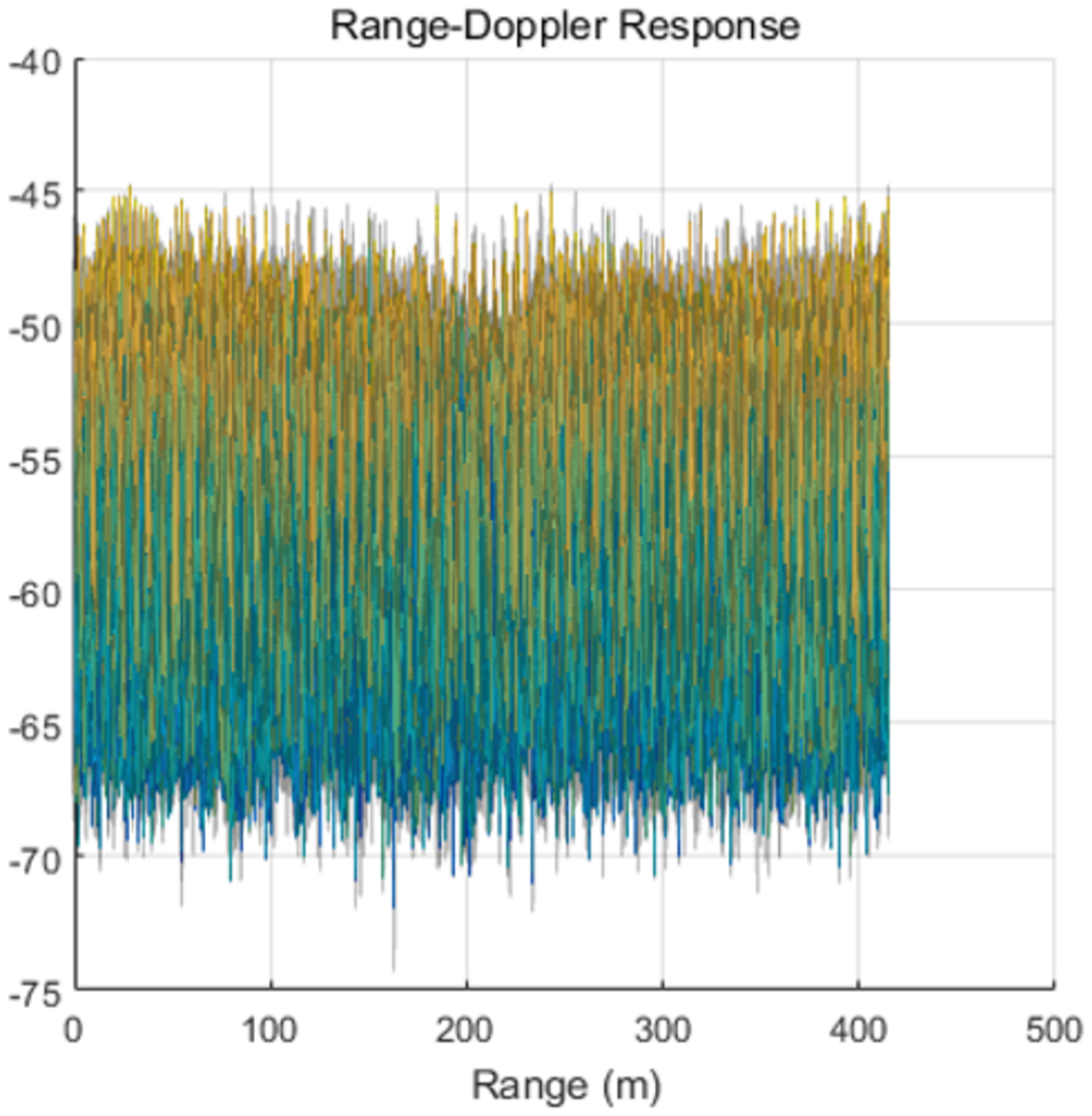} 
    \caption{No processing} 
    \label{fig7:d} 
    \vspace{4ex}
    \end{subfigure}%
  \caption{Simulated power levels with respect to range. Two targets exist in range 100m, 120m. Four interferences exist in range 40m, 50m, 60m, and 70m. Red circles are detected targets.}
  \label{fig7}
\end{figure}

We analyzed the interference mitigation performance of the proposed method. The results are shown in Fig. \ref{fig6}.
We use Fig. \ref{fig6}(\subref{fig6:a}) as deep learning label (not interfered), Fig. \ref{fig6}(\subref{fig6:b}) as deep learning input (interfered), and the deep learning output is Fig. \ref{fig6}(\subref{fig6:c}). We can see that the proposed deep learning algorithm finds out where the interference is. Under the considered situation, the reconstruction of the original signal is not perfect. However, we can see that the result of FFT in Fig. \ref{fig6}(\subref{fig6:f}) finds the object more clearly than the interfered input Fig. \ref{fig6}(\subref{fig6:e}). To compare result with other methods, we use the average signal to remaining interference noise ratio (SRINR)\cite{wagner2018threshold}. The SRINR result is in Table \ref{result}. Method \Rmnum{1} is time domain thresholding (TDT) method used in \cite{watanabe2007interference}. Method \Rmnum{2} did not use an adaptive threshold, which was proposed in \cite{wagner2018threshold}. The simulation SRINR is average of 50 random scenarios SRINR. Our proposed deep learning algorithm outperforms other methods. Especially, even in situations where the interference signal sources are close and the targets are too far away, our proposed method finds the target properly as shown in Fig. \ref{fig7}. 

\begin{table}[ht]
\caption{Simulation results}
\label{result}
\begin{center}
\begin{tabular}{c|c|c|c}
\hline
&Method \Rmnum{1}&Method \Rmnum{2}&Proposed\\
\hline
SRINR&23.369&22.665&26.091 \\
\hline

\end{tabular}
\end{center}
\end{table}

\section{Conclusion}
\label{sec_con}
In this paper, we proposed a novel approach to mitigate interference in CS radar system. We used a deep learning approach to mitigate interference. Our method shows better performance compared to other signal processing methods. Our method also shows good performance even when the target is far away. It is believed this method can be applied not only to CS waveforms but also to most situations where frequency changes linearly. This is because interference occurs at the point where the transmit signal crosses the interference signal. The interference patterns of linear frequency signals are similar. Experiments with other waveforms are left as future work. 
\section{ACKNOWLEDGEMENT}
This work is in part supported by Basic Science Research Program (NRF-2017R1A2B2007102) through NRF funded by
MSIP, Technology Innovation Program (10051928) funded by MOTIE, Bio-Mimetic Robot Research Center funded by DAPA
(UD130070ID), INMAC, and BK21-plus.

\bibliographystyle{IEEEtran}
\bibliography{my}

\end{document}